\documentclass[a4paper]{article}
\usepackage{hyperref,amssymb,amsmath,latexsym,amsthm,stmaryrd}
\usepackage{fullpage}
\usepackage{times}

\begin{document}
\title{Memory Undone: Between Knowing and Not Knowing in Data Systems}

\author{
  \begin{tabular}{cc}
    \textbf{Viktoriia Makovska} & \textbf{George Fletcher} \\
    Ukrainian Catholic University & Eindhoven University of Technology \\
    \href{mailto:makovska.pn@ucu.edu.ua}{makovska.pn@ucu.edu.ua} &
    \href{mailto:g.h.l.fletcher@tue.nl}{g.h.l.fletcher@tue.nl} \\
    & \href{https://gfletche.win.tue.nl}{gfletche.win.tue.nl} \\
    \\
    \textbf{Julia Stoyanovich} & \textbf{Tetiana Zakharchenko} \\
    New York University & Ukrainian Catholic University \\
    \href{mailto:stoyanovich@nyu.edu}{stoyanovich@nyu.edu} &
    \href{mailto:t.herasymova@ucu.edu.ua}{t.herasymova@ucu.edu.ua} \\
    \href{https://stoyanovich.org}{stoyanovich.org}
  \end{tabular}
}

\date{}
\maketitle
\thispagestyle{empty}

\paragraph{Memory in machine learning and data systems.}
Machine learning systems increasingly operate as infrastructures of memory. They collect, store, and learn from various datasets, many of which contain traces of personal, political, or cultural life. This raises a question: while data systems are designed for remembering, what role does a capability to forget play? The European Union's General Data Protection Regulation (GDPR) introduced the ``right to be forgotten,'' granting individuals the legal ability to request deletion of their personal data.\footnote{\url{https://eur-lex.europa.eu/eli/reg/2016/679/oj}} Permanently removing information from large-scale systems is technically challenging, since deletion must propagate through copies, caches, and derived data products, and most infrastructures were originally designed to accumulate rather than erase data \cite{data_protection_regulations}. Moreover, deleting records from database storage does not remove their influence on models trained on that data. In modern deep learning systems, once information is integrated into the weights of a neural network, its traces remain entangled in billions of parameters. This mismatch between legal frameworks and technical capabilities has made machine unlearning an urgent research problem. But the challenge goes beyond deletion. We need mechanisms to evaluate whether data traces persist or have been effectively unlearned inside these systems.
As an analogy, in quantum computation the quantum no-deleting theorem shows that an unknown quantum state cannot be deleted against a copy \cite{pati2000nodeleting}. Likewise, in machine learning, ``deletion'' is often not a complete removal, but of redistributing or attenuating influence across representations. Even after erasure or unlearning interventions, traces of prior data may persist indirectly, through learned correlations, semantic dependencies, or downstream artifacts. 

To reduce ambiguity, we distinguish several related practices.
\emph{Erasure} is an operational act that removes or disables access to data artifacts (e.g., rows, files, indexes, caches) \cite{koops2011forgotten}.
\emph{Unlearning} is an intentional technical intervention aimed at removing (or bounding) the influence of specific data on learned parameters and downstream outputs \cite{mu_original_def}.
\emph{Exclusion} refers to upstream decisions not to collect, label, or include certain people, events, or categories in datasets; it is a form of ``forgetting'' by omission \cite{data_feminism, onuoha}.
\emph{Forgetting}, in contrast, is an umbrella term for sociotechnical processes through which something becomes unavailable, varying in agency (intentional vs.\ incidental), temporality (momentary vs.\ durable), reversibility (recoverable vs.\ irreversible), and scale (individual records vs.\ population-level) \cite{seven_types_forgetting}.
On this view, both GDPR-backed deletion and the systematic non-inclusion training data are forms of ``forgetting,'' yet they operate through different mechanisms and raise distinct questions of governance.

Technical advances in unlearning demonstrate partial progress: exact unlearning in recommender systems \cite{ forget_me_now}, class forgetting without gradients \cite{kodge2024deep}, erasure in semantic web and description logic ontologies \cite{ChirkovaF09, Giacomo08} and selective unlearning in large language models \cite{wang2025selective}. Still, fundamental questions remain unresolved. What does it mean for a system to truly forget? How can erasure be ensured in the presence of semantic dependencies, where one record influences others? And at what point does forgetting protect rights, and when may it instead threaten them? For example, forgetting biased or harmful data may reduce harm, but forgetting underrepresented perspectives risks reproducing epistemic injustice \cite{muller2024data, muller2022}.

Forgetting is not an anomaly in data science but a consistent practice. At every stage of the pipeline, from data collection to feature design, to labeling, choices are made about what to exclude, ignore, or treat as noise. These acts of exclusion represent forms of forgetting. Prior research shows that data never exist in a neutral form: they become `data' only when transformed by human decisions shaped by social, political, and economic priorities \cite{sorting,data_ing_un_data_ing}. This opens the possibility of data silences \cite{onuoha}, or even systematic erasures where certain groups or perspectives are excluded altogether \cite{azoulay,caswell2021urgent}. At the infrastructure level, preservation is often treated as the default (``keep everything''), while deletion and unlearning appear as exceptional interventions motivated by rights (privacy) or necessity (harm reduction).

\paragraph{The balancing of knowing and not knowing.}
Work in participatory data modeling further underscores the critical balances between knowing and not knowing. Studies of entity–relationship design have revealed the presence of invisible entities---objects, actors, and relations that shape data models without being explicitly represented \cite{onion}. Designers often leave such elements undescribed to reduce complexity or maintain convenience. Their absence is not incidental but engineered, and it mirrors broader forgetting practices in data science. What is excluded can structure outcomes as powerfully as what is included. Similarly, data infrastructures contain nonrandom exclusions: the absence of data about certain populations comes from systematic marginalization or perceived lack of economic or political value \cite{lerman2013bigdata}. Forgetting, in this sense, is never neutral but an act of governance.

The stakes become clearer when considering information manipulation at scale. Research on Russian disinformation networks demonstrates how propaganda content, once statistically dominant in training corpora, can distort model outputs \cite{lies_in_chatbots}. Large language models reproduce such narratives, sometimes without disclaimers, illustrating a form of epistemic poisoning. 
Here, forgetting as a tool is not only about privacy but about resisting the integration of manipulative content into the ``memory'' of AI systems. Whether unlearning can provide remedies for such epistemic distortions remains an open question. What is clear, however, is that forgetting must be understood as both a technical safeguard and a political practice.

Modern epistemic infrastructures tend to centralize knowledge, reinforcing popular or canonical ideas and narrow the range of what is considered to be stored and what is erased. Epistemic decentralization, by contrast, enables participation in knowledge production from historically excluded actors. It can bring up unseen or shadowed perspectives, challenge existing assumptions, and correct what has been conventionally accepted as objective knowledge \cite{RodriguezMedina2025-RODDKE-2}. Francis Bacon warned through the Idols of the Theater, dominant intellectual frameworks can hide alternative ways of knowing and limit our possibilities to be open to unknown knowledge \cite{Bacon2000}.

\paragraph{Towards memory as a sociotechnical practice.}
In our work-in-progress we are investigating memory 
from a sociotechnical perspective, towards more socially responsive and just data and machine learning systems. Forgetting can be beneficial: it enables focus, reduces complexity, and removes harmful influence \cite{data_ing_un_data_ing}. Yet it can also marginalize, silence, or overwrite. Researchers of memory remind us that remembering and forgetting are intersected, to understand one requires acknowledging the other \cite{azoulay,bowker2008memory, cadena}. This framing is aligned with work in social sciences and memory studies that treats remembering/forgetting as institutionally organized practices, not merely individual cognitive events \cite{mesr_issue2}. It also resonates with research in cognitive psychology showing that human memory is reconstructive and susceptible to post-event framing effects \cite{loftus1974reconstruction}, and with evidence that post-decision attitude change can occur even without explicit memory for the decision (a phenomenon sometimes discussed as ``synthetic'' preference or satisfaction) \cite{lieberman2001amnesics}. In data systems, the database is often imagined as a fortress \cite{monson}, designed to protect and preserve memory. But memory is naturally selective, it becomes effective when it retains certain information while skipping others. Forgetting is not an exception to knowledge, but its condition. To know is always to begin from a position of not knowing \cite{caswell2021urgent,feinberg}. Knowing itself is never neutral: it assumes the power to select and keep authority over what is known, establishing a hierarchical relation between knower and known. In this context, not knowing can operate as a productive epistemic tool instead of being considered as a failure to know \cite{cadena}.

This perspective reframes machine unlearning not simply as a compliance tool, but as an opening to reimagine what knowing means in computational systems. If large language models are machines of remembering (aggregating vast corpora into probabilistic patterns) then forgetting must become a first-class capability.
At the same time, critical media theory argues that technology functions primarily as a retentional device, and is therefore structurally misaligned with ``forgetting'' \cite{stiegler2010memory}; we take this as a challenge to clarify what kinds of forgetting are possible (and which are not) in computational infrastructures.
This suggests developing models that are capable of unlearning by design, where forgetting is intentional, transparent, and accountable. Rather than treating forgetting as a limitation, unlearning can become a principle for healthier knowledge infrastructures.

Machine unlearning illustrates the fragile boundary: erasure can protect individuals, but it can also function as a tool of silencing. It highlights the dual role of forgetting and unlearning as both safeguard and risk. By reframing unlearning as a foundational step in the machine learning pipeline, this work aims to contribute to thinking about knowledge infrastructures: forgetting must be considered not as failure, but as a necessary, designed, and acknowledged practice.

\paragraph*{Goals of the talk.}
We propose to view forgetting in AI as a sociotechnical practice rather than a purely technical feature. This involves analyzing how unlearning operates within structures of power, rights, epistemic justice, and governance, and how it differs from forgetting and not knowing. Crucially, we treat forgetting/unlearning not as an internal optimization goal, but as a sociopolitical intervention with consequences beyond the data system---for example for privacy and redress, for the amplification or suppression of narratives, and for the potential silencing of underrepresented perspectives.

We also point toward the development of forgetting machines --- computational systems capable not only of remembering but of forgetting responsibly by design. By bringing together advances in machine unlearning with critical insights from philosophy, ethics, memory studies, and human–computer interaction, we argue that forgetting should be repositioned as an intentional and accountable practice in the design of future data systems.

Our aims in this presentation are as follows: (1) to present the reframing of forgetting in data systems as a sociotechnical practice; (2) to illustrate how acts of unlearning intersect with power, rights, epistemic justice, and governance; and (3) to open a broader discussion about how computational and epistemic infrastructures might be designed to forget responsibly by design. We aim to engage the conference participants in critical dialogue, sharing ideas and perspectives and building new connections and networks of scholars in Computer Science and beyond.

\paragraph*{Biographies.}
~
\smallskip

\noindent {\bf Viktoriia Makovska.} PhD researcher at the Ukrainian Catholic University (UCU), working on responsible data practices and machine unlearning. She holds a master's in data science from UCU and previously spent eight years in industry as a software engineer. She teaches decision support systems at the American University Kyiv and natural language processing at UCU.

\smallskip

\noindent {\bf George Fletcher.} George studies data management systems with a particular focus on the social and human dimensions of query and schema languages. He has been on the faculty at Eindhoven University of Technology for the past 15 years and serves on the boards of the EDBT Association, ACM Transactions on Database Systems, ACM SIGMOD Record, and Transactions on Graph Data and Knowledge. He is a regular program committee member for EDBT, VLDB, ICDE, and SIGMOD, and has recently co-organized interdisciplinary Shonan and NIAS Lorentz Center seminars that bring together computer scientists, humanities scholars, and social scientists to explore algorithms and data as sociotechnical interventions. He holds a PhD in Computer Science from Indiana University Bloomington and completed undergraduate studies in mathematics and cognitive science at the University of North Florida.

\smallskip

\noindent {\bf Julia Stoyanovich.} Dr. Julia Stoyanovich is the Institute Associate Professor of Computer Science and Engineering, Associate Professor of Data Science, and Director of the Center for Responsible AI at New York University. Her mission is to make ``Responsible AI'' synonymous with ``AI.'' She pursues this goal through academic research, education, technology policy, and public engagement, regularly speaking about both the benefits and the risks of AI. Her research spans data management and AI systems, as well as the ethics and governance of AI. Julia holds an M.S. and Ph.D. in Computer Science from Columbia University, and a B.S. in Computer Science and in Mathematics \& Statistics from the University of Massachusetts Amherst. She is a recipient of the Presidential Early Career Award for Scientists and Engineers (PECASE) and is a Senior Member of the Association for Computing Machinery (ACM).

\smallskip

\noindent {\bf Tetiana Zakharchenko.}
Tetiana is the Director of the Computer Science Program and the Center for Data, AI, and Society at the Ukrainian Catholic University. Her work lies at the intersection of technology and society, with a focus on responsible AI, algorithmic fairness, and data science in education. Tetiana holds a Ph.D. in Mathematics from Taras Shevchenko National University of Kyiv and a MicroMasters in Statistics and Data Science from MIT. She has been a Research Fellow at the Center for Responsible AI at New York University. Tetiana is a member of the ELLIS Community and serves on the Committee on Artificial Intelligence Development of the Ministry of Digital Transformation of Ukraine.

\bibliographystyle{plain} 
\bibliography{refs}
\end{document}